\documentclass{article}
\usepackage{amsmath}
\usepackage{graphicx}
\usepackage{amsfonts}
\usepackage{amssymb}
\usepackage{geometry}
\usepackage{float}
\usepackage{bm}
\usepackage{color}
\usepackage{cite}
\usepackage{subfigure}
\usepackage{geometry}
\usepackage{geometry}

\begin{document}

\title{The condensation of ideal Bose gas in a gravitational field in the
framework of Dunkl-statistic }
\author{B. Hamil\thanks{%
hamilbilel@gmail.com } \\
D\'{e}partement de Physique, Facult\'{e} des Sciences Exactes, \\
Universit\'{e} de Constantine 1, Constantine, Algeria \and B. C. L\"{u}tf%
\"{u}o\u{g}lu\thanks{%
bekir.lutfuoglu@uhk.cz (corresponding author)} \\
Department of Physics, University of Hradec Kr\'{a}lov\'{e}, \\
Rokitansk\'{e}ho 62, 500 03 Hradec Kr\'{a}lov\'{e}, Czechia.}
\date{}
\maketitle

\begin{abstract}
In the framework of the theory of Dunkl-deformed bosons, Bose-Einstein condensation of two and three-dimensional Dunkl-boson gases confined in the 
one-dimensional gravitational field is investigated. Using the semi-classical
approximation method, we calculate the expressions of the Dunkl-critical
temperature $T_{c}^{D}$, the ground state population $\frac{N_{0}^{D}}{N}$
and the Dunkl-mean energy and Dunkl-specific heat functions. Further numerical calculation shows that the
condensation temperature ratio $\frac{T_{c}^{D}}{T_{c}^{B}}$ increases with the
increasing Wigner parameter.
\end{abstract}

\section{Introduction}
Exactly a century ago, Einstein predicted a {process in a} state of matter, known as the Bose-Einstein condensation (BEC) \cite{Einstein}, by evaluating the quantum formulation given by Bose \cite{Bose} via a personal letter to him. According to this estimation, a finite fraction of the number of particles should start to condense into their lowest-energy states below a particular temperature. After many decades this quantum-statistical phase transition is  observed experimentally in the laboratory firstly by Cornell, Wieman, and then by Ketterle for the rubidium  \cite{Cornell} and sodium atoms  \cite{Ketterle}, respectively. The 2001 Nobel Prize in Physics was jointly awarded to these physicists for their studies which expanded our understanding at the quantum scale and led to the observation of new physical effects. Since then, an increasing number of studies have dealt with the BEC, both experimentally and theoretically \cite{Grossmann, Napolitano, Pessoa, Schneider, Dalfovo, Anglin, LiuWu, Wu, Alikawa, Gaunt, Zapf, Levinsen, Wachtler, Levkov, Cabrera, Gerbier, Roccuzzo, Drescher, Yao, Xie}.

The role of external potential on BEC is widely discussed in the literature. For example, Bagnato et al. employed a generic power-law potential energy form,  and they estimated the critical temperature and ground state population of an ideal Boson gas in three dimensions \cite{Bagnato87}. Later Kirsten et al. considered spin-0 particles system and analyzed the BEC with isotropic and anisotropic harmonic potential traps \cite{Kirsten96}. The BEC of non-relativistic systems under the isotropic harmonic oscillator potential were examined in one and three dimensions by the Ketterle et al. \cite{Ketterle}, and in two dimensions by Mullin \cite{Mullin}. In literature, we observe many works that took into account the uniform gravitational field effects. For instance, Gersch explored the thermodynamics of the Bose-Einstein gas in the presence of gravitational field \cite{Gersch}. Widom demonstrated theoretically the condensation of the ideal Bose liquid  which is confined in the gravitational field \cite{Widom}. Baranov et al. studied the two-dimensional BEC of atoms trapped in a rectangular well under the influence of the gravitational field \cite{Baranov}. Rivas et al. obtained the condensation temperature of the particle system for two different trapping cases and discussed the modifications in the derived temperature taking into account the homogeneous gravitational field \cite{Rivas}.  Liu et al. considered a non-interacting Bose gas system in the presence of a uniform gravitational field in one dimension and they derived the condensation temperature and condensate fraction within the semi-classical approach \cite{Liu}. Later, Du et al. revisited the same problem within two and three dimensions \cite{Du}, and presented the extended results of \cite{Liu}.

On the other hand, in the middle of the last century, Wigner started an interesting discussion: "Instead of obtaining the equation of motions from the commutation relations, can we derive the corresponding commutation relations from the equation of motions?" \cite{Wigner}. To answer this question, Wigner dealt with the free case and the classical harmonic oscillator problems, and he showed that the commutation relations of these two problems can be obtained with a constant difference. Since the constant can take arbitrary values, he concluded that the reverse case does not provide a unique answer. Just a year later, Yang handled the same question with the quantum harmonic oscillator problem \cite{Yang}. He showed that arbitrariness vanishes if a strict definition of Hilbert space is taken into account with a rigorous series expansion. However, during his proof, he had to introduce a reflection operator, $\hat{R}$, where $\hat{R}\psi(x)=\psi(-x)$, by deforming the momentum operator in the spatial space. Several decades later, mathematicians were investigating the relations between differential-difference and reflection operators. For this context, Dunkl presented the Dunkl derivative of the form \cite{Dunkl1}
\begin{eqnarray}
    D_x&=& \frac{d}{dx}+ \frac{{\theta}}{x} \left(1-\hat{R}\right), \label{cd}
\end{eqnarray}
which slightly differs from Yang's derivative by the second term of Eq. \eqref{cd}. {Because of this historical connection, $\theta$ is named the Wigner parameter.  Basically, this real-valued parameter determines the parity effects and it has a lower bound given by $\theta>-1/2$.} The Dunkl derivative aroused the interest of physicists as well as mathematicians. At first, this interest was manifested in the use of the Dunk derivative in the Calogero-Sutherland-Moser model \cite{Lapointe, Kakei}.
In the last decade, research with Dunkl-formalism has spread to the broad field of physics. For example, in relativistic and non-relativistic quantum mechanics context two and three dimensional isotropic and anisotropic Dunkl-oscillators are investigated in \cite{G1, G2, G3, G4}. The relativistic oscillators, namely the Dunkl-Dirac oscillator are examined in \cite{Sargol, Mota1, Bilel3}, while the Dunkl-Klein-Gordon oscillators are studied in \cite{Bilel3, Mota2, Mota3, Bilel1}. Similarly, the partial solution of the  Dunkl-Duffin-Kemmer-Petiau oscillator is explored in \cite{Merad}.  Dunkl-Newton mechanics, Dunkl-electrostatics, and Dunkl-Maxwell equations are constructed in \cite{Chungrev, ChungDunkl}, respectively. Recent studies in Dunkl-general relativity
and Dunkl-black hole thermodynamics are given in \cite{DRelat1, DRelat2}. Dunkl- Boson statistic mechanics are formed  in \cite{Marcelo}. The dynamics of electrons in a graphene layer which is under the effect of {an} external magnetic field is derived in  \cite{Bilel2}. There are many more studies in the literature that takes into account the Dunkl formalism as it allows itself to discuss parity-dependent solutions simultaneously \cite{G5,  Jang, Ramirez1, Ramirez2, Chung1, Ghaz, Kim, Ojeda, ChungEPJP, Hassan, Dong, Seda, Mota2022, new1, new3, new4, quezada}.

Very recently, we investigated the statistical mechanics of Dunk-ideal Bose gas and Dunk-blackbody radiation in \cite{1}. Since gravitational field effects are very crucial for the BEC, in this manuscript we intend to explore the gravitational field effects on condensation with {a} semi-classical method. In particular, we consider two and three-dimensional ideal Bose-gas and derive its critical temperature, condensation rate, and specific heat in the Dunkl formalism. To this end, we construct the manuscript as follows: In sections 2 and 3, we study three and two-dimensional Dunkl-Bose gases, respectively. In the next section, we discuss the findings graphically and conclude the manuscript.

\section{Three-dimensional system}

Let us consider a system consisting of $N$ bosons confined in the gravitational field along $x$ direction, which can be expressed as
\begin{equation}
V\left( x\right) =\left \{ 
\begin{array}{c}
mgx\text{ \  \  \  \ (}x>0\text{)} \\ 
\infty \text{ \  \  \  \  \  \  \  \ (}x\leq 0\text{)}%
\end{array}%
\right. ,  \label{1}
\end{equation}
Here, $m$ is the mass of a boson particle, and $g$ is the acceleration of gravity. In the grand canonical Dunkl-statistics the total number of bosons, $N$, reads \cite{1}
\begin{equation}
N=N_{0}^{D}+N_{e}^{D},  \label{n}
\end{equation}
where%
\begin{equation}
N_{0}^{D}=\frac{2}{z^{-2}-1}+\frac{(1+2\theta )}{z^{-(1+2\theta )}+1},
\label{N0}
\end{equation}%
stands for the number of condensed particles, and  
\begin{equation}
N_{e}^{D}=\sum_{i\neq 0}\frac{2}{e^{2\beta E_{i}}z^{-2}-1}+\sum_{i\neq 0}%
\frac{(1+2\theta )}{e^{\beta (1+2\theta )E_{i}}z^{-(1+2\theta )}+1},
\label{tot}
\end{equation}
denotes the particles in excited states. Here, $\theta$ 
indicates the Wigner parameter, $z=e^{\beta \mu }$ corresponds to the fugacity, where $\mu $ symbolizes the chemical potential of a boson particle, and $\beta =\frac{1}{KT}$ is the inverse temperature with the Boltzmann's constant, $K$, and the temperature of the system, $T$. For very large numbers of particles, it is hard to evaluate these sums analytically. One of the most appropriate methods for performing this analysis is to change the sum with ordinary integrals weighted by a proper  density of states, $\rho \left( E\right) $. Within this approach Eq. (\ref{tot}) becomes%
\begin{equation}
N_{e}^{D}=2\int_{0}^{+\infty }\frac{\rho \left( E\right) dE}{e^{2\beta
E}z^{-2}-1}+(1+2\theta )\int_{0}^{+\infty }\frac{\rho \left( E\right) dE}{%
e^{\beta (1+2\theta )E}z^{-(1+2\theta )}+1}.  \label{EX}
\end{equation}
In the presence of a gravitational field, the density of states changes and the thermal behavior differs from that of free gas. In the semi-classical approach, we can consider each quantum state to exist in a phase space cell with volume $\Omega =h^{3}$. Then, we calculate the three-dimensional energy density of states by the standard method with
\begin{equation}
\rho \left( E\right) =\frac{1}{h^{3}}\int \int d\overrightarrow{r}d%
\overrightarrow{p}\delta \left( H-E\right) =\frac{\sqrt{2m}L^{2}}{3g\pi
^{2}\hbar }E^{3/2}.  \label{DE}
\end{equation}%
By utilizing Eq. (\ref{DE}), we transform Eq. (\ref{EX}) into 
\begin{equation}
N_{e}^{D}=\frac{L^{2}\sqrt{\pi m}}{8\pi ^{2}g\hbar ^{3}}\left( K_{B}T\right)
^{5/2}g_{5/2}{\left( \theta, z\right)} .  \label{ca}
\end{equation}
Here, we introduce the generalized Bose functions, $g_{s}(z,\theta )$, which are defined by \cite{1}
\begin{equation}
g_{s}\left( \theta ,z\right) =g_{s}(z^{2})-\left( \frac{2}{1+2\theta }%
\right) ^{s-1}g_{s}(-z^{1+2\theta }),
\end{equation}%
$\allowbreak $
where the Bose functions, $g_{s}(z)$, are given with%
\begin{equation}
g_{s}(z)=\frac{1}{\Gamma \left( s\right) }\int_{0}^{+\infty }\frac{x^{s-1}}{%
e^{x}z^{-1}-1}dx.
\end{equation}
Similar to the ordinary case, the BEC of a trapped Dunkl-boson system with a finite number of particles should not have an evident critical temperature {\cite{Liu, Du, Pessoa, bilokon}}. However, one still can determine an effective Dunkl-critical temperature, $T_{c}^{D}$, via two methods \cite{Du}:
\begin{itemize}
    \item By taking the chemical potential as zero, and assuming all particles are in the excited states.
    \item By considering the temperature at which the specific heat of the system reaches its maximum value.  
\end{itemize}
Here, we prefer to use the first method because of its simplicity. After straightforward manipulations, we obtain the Dunkl-BEC temperature of the ideal Bose gas.
\begin{equation}
K_{B}T_{c}^{D}=\left( \frac{32\pi g\hbar ^{3}N}{3L^{2}\sqrt{m}\zeta \left( 
\frac{5}{2}\right) }\right) ^{2/5}\left[ 1+\frac{2\sqrt{2}-1}{\left(
1+2\theta \right) ^{3/2}}\right] ^{-2/5}.  \label{D}
\end{equation}%
Here, $\zeta(n)$ denotes the Riemann-Zeta function. It is worth noting that the Dunkl-critical temperature depends not only on the total number of particles $N$, the length of the system $L$, {and} the mass of the particles $m$ but also on the Wigner parameter. If we consider the limit, $\theta \rightarrow 0$, then the Dunkl-critical temperature reduces to the ordinary Bose critical temperature, $T_{c}^{B}$, given in the form of
\begin{equation}
K_{B}T_{c}^{B}=\left( \frac{16\pi g\hbar ^{3}N}{3L^{2}\sqrt{2m}\zeta \left( 
\frac{5}{2}\right) }\right) ^{2/5}.  \label{bose}
\end{equation}
By matching Eq. \eqref{D} and Eq. \eqref{bose}, we express the ratio
\begin{equation}
\frac{T_{c}^{D}}{T_{c}^{B}}=\left( \frac{2\sqrt{2}}{1+\frac{2\sqrt{2}-1}{%
\left( 1+2\theta \right) ^{3/2}}}\right) ^{2/5}.  \label{TC}
\end{equation}%
Now, let us to investigate the condensate fraction. For $T\leq T_{c}^{D}$, we first express the ground state population, $\frac{N_{0}^{D}}{N}$, 
according to Eq. (\ref{n}) and Eq. (\ref{ca})
\begin{equation}
\frac{N_{0}^{D}}{N}=1-\left( \frac{T}{T_{c}^{D}}\right) ^{5/2}.
\end{equation}
Then, by using Eq. (\ref{TC}), we obtain the ground state population for three dimensional system as follows:%
\begin{equation}
\frac{N_{0}^{D}}{N}=1-\frac{1}{2\sqrt{2}}\left( 1+\frac{2\sqrt{2}-1}{\left(
1+2\theta \right) ^{3/2}}\right) \left( \frac{T}{T_{c}^{B}}\right) ^{5/2}.
\end{equation}
We find it interesting to investigate the thermodynamic quantities of ideal Bose gas in the presence of a gravitational field in the framework of the Dunkl-statistics. At first, we focus on the internal energy, $U$, according to the following formula
\begin{equation}
U=\sum_{i}N_{i}E_{i}.  \label{K}
\end{equation}
After converting the sum into an integral, we obtain the following expression for the
internal energy.%
\begin{equation}
U=\frac{5L^{2}\sqrt{\pi m}}{32\pi ^{2}g\hbar ^{3}}\left( K_{B}T\right)
^{7/2}g_{7/2}(\theta ,z).  \label{int}
\end{equation}%
For the case $T\leq T_{c}^{D}$, one can set $z=1$ and utilize Eq. (\ref{int}) to obtain the internal energy of the system directly. However, for the case $T>T_{c}^{D}$, first we have to determine the value of $z$ via Eqs. (\ref{n}), (\ref{N0}) and (\ref{ca}). Then, by substituting the value of $z$ in Eq. (\ref{int}), we can be able to express the internal energy function.

Next, we examine the specific heat function. For the condensate phase case, $T\leq
T_{c}^{D}$, we put $z=1$ and derive the thermal quantity via, $C=\left. \frac{\partial
U}{\partial T}\right \vert _{N,V}$,  as follows:
\begin{equation}
\frac{C_{<}}{K_{B}N}=\frac{35}{8}\frac{g_{7/2}(1)}{g_{5/2}(1)}\left[ \frac{1+%
\frac{4\sqrt{2}-1}{\left( 1+2\theta \right) ^{5/2}}}{1+\frac{2\sqrt{2}-1}{%
\left( 1+2\theta \right) ^{3/2}}}\right] \left( \frac{T}{T_{c}^{D}}\right)
^{5/2},
\end{equation}
or equivalently in terms of the usual condensate temperature as
\begin{equation}
\frac{C_{<}}{K_{B} N}=\frac{35}{16\sqrt{2}}\frac{g_{7/2}(1)}{g_{5/2}(1)}\left[
1+\frac{4\sqrt{2}-1}{\left( 1+2\theta \right) ^{5/2}}\right] \left( \frac{T}{%
T_{c}}\right) ^{5/2}.
\end{equation}%
For the gas phase case, $T>T_{c}^{D},$  we have $%
N_{0}^{D}=0$, and $z\neq 1$. By differentiate Eq. (\ref{int}), we obtain%
\begin{eqnarray}
\frac{d}{dT}U &=&\frac{7K_{B}}{2}\frac{5L^{2}\sqrt{\pi m}}{32\pi ^{2}g\hbar
^{3}}\left( K_{B}T\right) ^{5/2}\left[ g_{7/2}(z^{2})-\left( \frac{2}{%
1+2\theta }\right) ^{5/2}g_{7/2}(-z^{1+2\theta })\right]  \notag \\
&&+\frac{2}{z}\frac{dz}{dT}\frac{5L^{2}\sqrt{\pi m}}{32\pi ^{2}g\hbar ^{3}}%
\left( K_{B}T\right) ^{7/2}\left[ g_{5/2}(z^{2})+\left( \frac{2}{1+2\theta }%
\right) ^{3/2}g_{5/2}(-z^{1+2\theta })\right] ,  \label{du}
\end{eqnarray}%
where we employed the relation%
\begin{equation}
\frac{d}{dz}g_{s}(z^{n})=\frac{n}{z}g_{s-1}(z^{n}).
\end{equation}
The quantity $\frac{2}{z}\frac{dz}{dT}$ can be obtained by the total particle number $N$ that is given by Eq. (\ref{n}). Since it is a constant, we  have, $\frac{d}{dT}N=0$, which leads to
\begin{equation}
\frac{2}{z}\frac{dz}{dT}=-\frac{5K_{B}}{2}\frac{1}{\left( K_{B}T\right) }%
\left[ \frac{g_{5/2}(z^{2})-\left( \frac{2}{1+2\theta }\right)
^{3/2}g_{5/2}(-z^{1+2\theta })}{g_{3/2}(z^{2})+\left( \frac{2}{1+2\theta }%
\right) ^{1/2}g_{3/2}(-z^{1+2\theta })}\right] .
\end{equation}%
We substitute the above expression into Eq. (\ref{du}), and express the Dunkl-specific heat function of the gas phase as:%
\begin{equation}
\frac{C_{>}}{K_{B}N}=\frac{35}{8}\left \{ \frac{g_{7/2}(\theta ,z)}{%
g_{5/2}(\theta ,z)}-\frac{5}{7}\frac{g_{5/2}(\theta ,z)+2\left( \frac{2}{%
1+2\theta }\right) ^{3/2}g_{5/2}(-z^{1+2\theta })}{g_{3/2}(\theta
,z)+2\left( \frac{2}{1+2\theta }\right) ^{1/2}g_{3/2}(-z^{1+2\theta })}%
\right \} .
\end{equation}

\newpage
\section{Two-dimensional system}

In this section, we study the $d=2$ case with the methodology followed in the previous section.  We assume that the gravitational field is toward the $x$ direction, as it is given with Eq. (\ref{1}) in the three-dimensional case. Here, the density of state differs from Eq. \eqref{DE}, and appears in the following form:
\begin{equation}
\rho =\frac{LE}{2\pi g\hbar ^{2}}.  \label{2d}
\end{equation}
By using Eq. (\ref{2d}) in Eq. (\ref{EX}), we find the Dunkl-corrected number of particles in the excited states 
\begin{equation}
N_{e}^{D}=\frac{L}{4\pi g\hbar ^{2}}\left( K_{B}T\right) ^{2}g_{2}\left(
\theta ,z\right) .
\end{equation}
Then, we derive the Dunkl-critical temperature by a direct calculation as performed in the $d=3$ dimensions.%
\begin{equation}
K_{B}T_{c}^{D}=\frac{\pi ^{2}}{6}\sqrt{\frac{2\pi g\hbar ^{2}N}{L}}\left( 
\frac{1+2\theta }{1+\theta }\right) ^{1/2}.  \label{tx}
\end{equation}
In the limit of $\theta \rightarrow 0$, it is demoted to the ordinary Bose condensation temperature, 
\begin{equation}
K_{B}T_{c}^{B}=\frac{\pi ^{2}}{6}\sqrt{\frac{2\pi g\hbar ^{2}N}{L}}.
\end{equation}%
Following these considerations, we easily get the existing relation between the Dunkl-condensation temperature and the standard one:
\begin{equation}
\frac{T_{c}^{D}}{T_{c}^{B}}=\left( \frac{1+2\theta }{1+\theta }\right)
^{1/2}.  \label{ccc}
\end{equation}%
By combining Eqs. (\ref{tx}) and (\ref{n}), we get the Dunkl-corrected condensate fraction in the temperature region $T\leq T_{c}^{D},$ 
\begin{equation}
\frac{N_{0}^{D}}{N}=1-\left( \frac{T}{T_{c}^{D}}\right) ^{2},
\end{equation}
which can also be written as%
\begin{equation}
\frac{N_{0}^{D}}{N}=1-\left( \frac{1+\theta }{1+2\theta }\right) \left( 
\frac{T}{T_{c}^{B}}\right) ^{2}.
\end{equation}%
Analogous to the previous section, we finally derive Dunkl-corrected internal energy and specific heat functions. To this end, we use Eq. (\ref{K})  and convert the sum into an integral. After the algebra, we obtain the internal energy function in the form of
\begin{equation}
U=\frac{L}{4\pi g\hbar ^{2}}\left( K_{B}T\right) ^{3}g_{3}\left( \theta
,z\right) .
\end{equation}%
After that with the help of the  thermodynamic definition of specific heat function, we get
\begin{equation}
\frac{C}{K_{B}N}=\left \{ 
\begin{array}{ll}
\frac{3}{2}\left( \frac{T}{T_{c}^{B}}\right) ^{2}\frac{g_{3}(1)}{g_{2}(1)}%
\left[ 1+\frac{3}{(1+2\theta )^{2}}\right] &\text{ \  \  \ for }T\leq T_{c}^{D}
\\ 
3\left[ \frac{g_{3}(\theta ,z)}{g_{2}(\theta ,z)}\right] -2\left[ \frac{%
g_{2}(\theta ,z)+\frac{4}{(1+2\theta )}g_{2}(-z^{1+2\theta })}{g_{1}(\theta ,z)+2g_{1}(-z^{1+2\theta })}\right] &\text{ \  \  \ for }T\geq T_{c}^{D}%
\end{array}%
\right. 
\end{equation}

\section{Discussions and Conclusion}

\begin{figure}[htbp]
\centering
\includegraphics[scale=0.85]{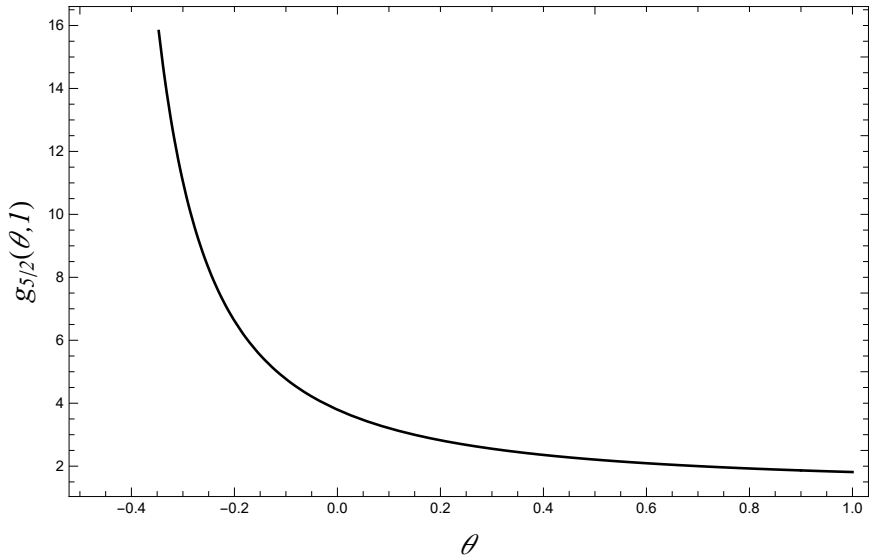}
\caption{The variation of the function $g_{5/2}\left( \protect \theta %
,1\right) $ versus the Wigner parameter.} \label{Fig1}
\end{figure}

Now, we can display our results graphically. For this, we first need to show the behavior of the generalized Bose functions. In Fig. \ref{Fig1}, we depict $g_{5/2}(\theta,1)$ function versus the Wigner parameter. 
 
\begin{figure}[htbp]
\centering
\includegraphics[scale=0.85]{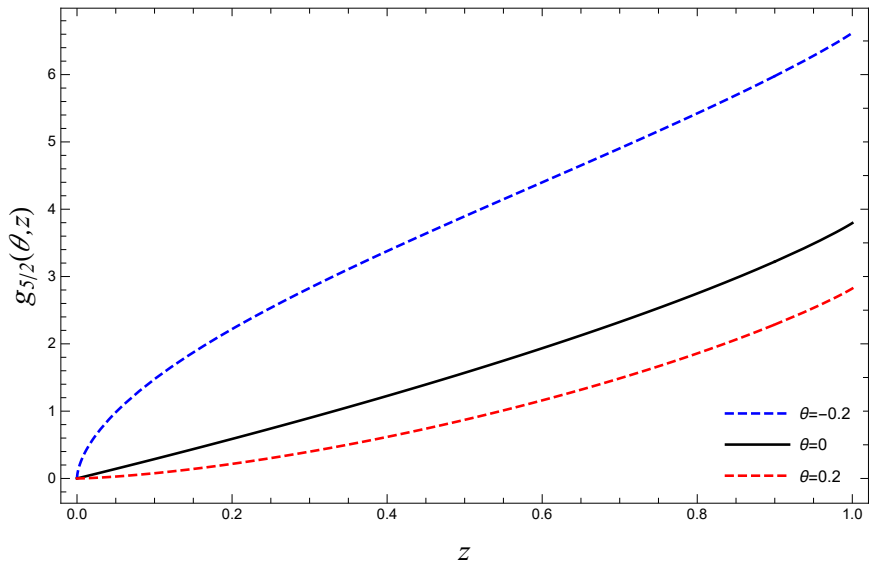}
\caption{The variation of the function $g_{5/2}\left( \protect \theta,z\right) $ versus $z$ for different values of Wigner parameter.} \label{Fig2}
\end{figure}

\newpage We observe that the generalized Bose function decreases monotonically for increasing values of the Wigner parameter. Then, we plot  $g_{5/2}(\theta,z)$ function versus $z$ in Fig. \ref{Fig2}.

\begin{figure}[tbph]
\centering \includegraphics[scale=0.85]{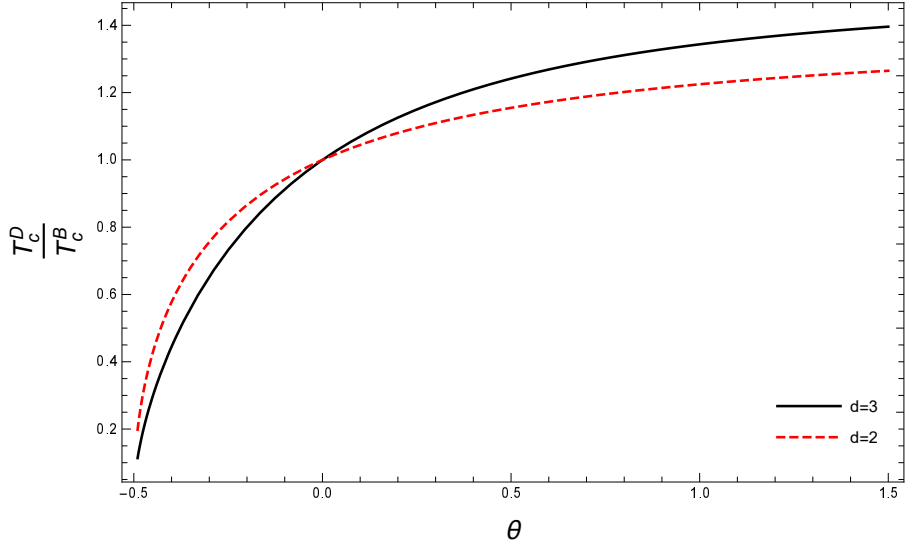}
\caption{The variation of $\frac{T_{c}^{D}}{T_{c}^{B}}$ as {the} function of the Wigner parameter in two and three dimensions.} \label{Fig3}
\end{figure}

We observe that Dunkl's correction increases the values of generalized Bose functions when the Wigner parameter takes negative values, and decreases it when it takes positive values. Next, in Fig. \ref{Fig3} we show the change of the condensation temperature ratio via the Wigner parameter in two and three dimensions.

\newpage

We observe that in both dimensions this rate increases monotonically with the increasing Wigner parameter. However, the value of {the} Dunkl-condensation temperature ratio in two dimensions is greater in the region where the Wigner parameter is negative than in three dimensions. In the interval where the Wigner parameter is positive, the rate in three {dimensions} becomes greater. Finally, we demonstrate the condensate fraction $\frac{N_{0}^{D}}{N}$ versus normalized temperature $\frac{T}{T_{c}^{B}}$ for different values of the Wigner parameter in two and three dimensions. 

\begin{figure}[htbp]
\centering \includegraphics[scale=0.85]{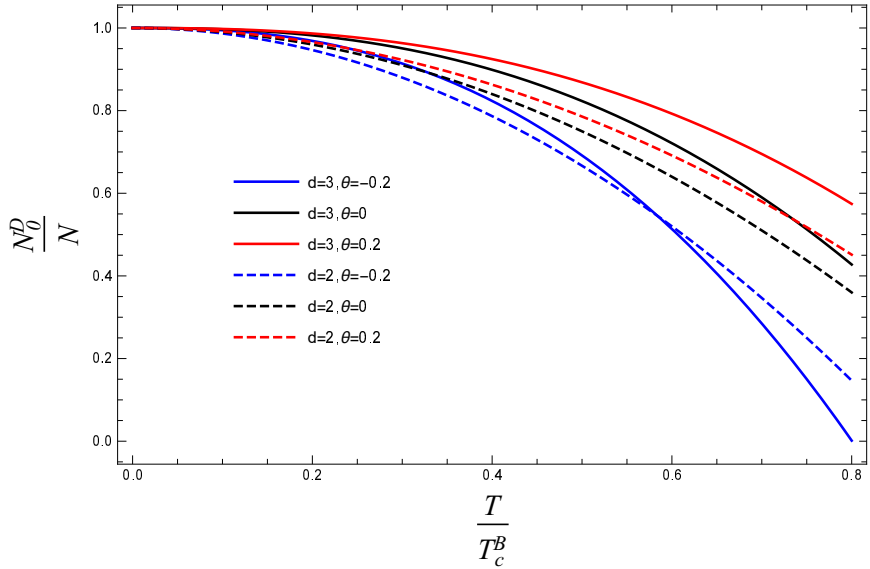}
\caption{The condensate fraction $\frac{N_{0}^{D}}{N}$ as {the} function $\frac{T}{T_{c}^{B}}$ for different values of $\protect \theta $ for $T\leq T_{c}^{D}$ in two and three dimensions.}
\end{figure}

Here, we see that solid lines (three-dimensional), no matter the Wigner parameter values, are greater than the dashed lines(two-dimensional) up to a ratio value of the normalized temperature. This ratio value depends on the Wigner parameter. This shows that the parity symmetry of the particles {is} important on the {BEC}.

In this manuscript, we consider the {BEC in} the presence of {the} gravitational field in two and three dimensions within the Dunkl-formalism. The semi-classical approach shows that the critical temperature increases with increasing  values of the Wigner parameter; so by using other Bose gases which have smaller Wigner parameter value{s}, we can obtain the Bose condensation at smaller temperatures.

\section*{Acknowledgments}
BCL is supported by the Internal  Project,  [2023/2211],  of  Excellent  Research  of  the  Faculty  of  Science  of Hradec Kr\'alov\'e University.

\section*{Data Availability Statements}
The authors declare that the data supporting the findings of this study are available within the article.

\end{document}